\documentclass[%
 reprint,onecolumn,
 amsmath,amssymb,
 aip, 12pt
]{revtex4-2}
\usepackage{xcolor}
\usepackage{graphicx}
\usepackage{dcolumn}
\usepackage{bm}
\usepackage{hyperref}
\usepackage{setspace}
\usepackage{threeparttablex}
\usepackage{booktabs}
\usepackage{array}
\usepackage[title]{appendix}
\newcommand{\bra}[1]{\ensuremath{\left\langle#1\right|}}
\newcommand{\ket}[1]{\ensuremath{\left|#1\right\rangle}}
\newcommand{\ketbra}[2]{\ensuremath{\left|#1\right\rangle\left\langle#2\right|}}

\newcolumntype{P}[1]{>{\centering\arraybackslash}p{#1}}
\begin{document}

\title{Raman spectroscopy at metal interfaces: A numerical study of the strong coupling regime}
\author{Zeyu Zhou}
\email{zz9884@princeton.edu}
\affiliation{Department of Chemistry, Princeton University, Princeton, NJ 08544, United States}
\author{Maxim Sukharev}
\email{Maxim.Sukharev@asu.edu}
\affiliation{Department of Physics, Arizona State University, Tempe, Arizona 85287, United States}
\affiliation{College of Integrative Sciences and Arts, Arizona State University, Mesa, Arizona 85212, United States}
\author{Abraham Nitzan}
\email{anitzan@sas.upenn.edu}
\affiliation{Department of Chemistry, University of Pennsylvania, 231 South 34th Street, Philadelphia, Pennsylvania 19104, United States}
\affiliation{School of Chemistry, The Raymond and Beverly Sackler Faculty of Exact Sciences and The Sackler Center for computational Molecular and Materials Science, Tel Aviv University, Tel Aviv 6997801, Israel}
\author{Joseph E. Subotnik}
\email{subotnik@princeton.edu}
\affiliation{Department of Chemistry, Princeton University, Princeton, NJ 08544, United States}

\date{\today}

\begin{abstract}
We investigate how proximity to a metal nanostructure, particularly to a flat mirror or a cavity confined between two mirrors, affects the vibronic structure of Raman scattering signals. We find that such proximity, particularly for the strong-coupling situation encountered in cavity environments, plays multiple roles in shaping Raman signals beyond the now-familiar signal enhancement known as surface-enhanced Raman scattering (SERS). First, in analogy to the electromagnetic SERS mechanism, near or between mirrors, the local field experienced by a molecule differ from that in vacuum. In particular, between mirrors, the cavity enhances the effective excited state population by trapping the EM field inside it. Second, the nearby metal surface provides a relaxation channel and a lineshape broadening mechanism, and  inside a cavity this lineshape is inherited by the cavity polaritons. 
This relaxation results in a loss of yield but the associated broadening also leads to significant absorption over a  larger frequency range. 
Third, near metallic interfaces interference between incident and reflected light can lead to a richly structured Raman spectrum. For instance,  we find that the Rabi contraction (that results from depopulating the ground state)  can interfere with Raman signals (and the effect appears to be the same order as Raman itself). These cavity effects are calculated by a full-scale FDTD simulation and highlight the convoluted but fascinating roles of photonic materials on optical signals.
\end{abstract}
\maketitle

\newpage
\section{Introduction}
The enhancement of light–matter interactions near metal and semiconductor interfaces have been a central research area for the last few decades. Scientists possess unprecedented tools with the development of nanotechnologies, such as metallic/composite dielectric nanolayers and nanoparticles. For example, Raman-active molecules near metallic surfaces enables surface-enhanced Raman spectroscopy (SERS), magnifying signals with rich vibrational signatures.\cite{haynes2005surface, stiles2008surface, frontiera2011surface, sharma2012sers, yi2025surface, boolakee2022light} 
Another fruitful area of research involves embedding materials in between two mirrors, leading to the formation of hybrid light–matter states within such nanostructures.\cite{bayer2001coupling, walther2006cavity, hutchison2012modifying, sweeney2014cavity, ebbesen2016hybrid, hummer2016cavity,saunders2016cavity,kavokin2017microcavities, zhang2018dicke,ribeiro2018polariton, frisk2019ultrastrong, hertzog2019strong, garcia2021manipulating, li2021cavity, hu2022tuning} 
The latter phenomena arise from enhanced effective coupling, driven by prolonged cavity photon lifetimes and reduced modal volumes and have led to many opportunities to engineering compact, highly responsive and high signal-to-noise ratio nano-optical devices with tailored functionalities for sensing, spectroscopy, and energy harvesting.
Because these devices are of nanometric scale by design, a proper analysis of their capabilities requires a comprehensive semiclassical method that can capture essential quantum effects and macroscopic electrodynamics interaction with heterogeneous optical structures. 

To this end, simulations combining classical electrodynamics—implemented through finite-difference time-domain (FDTD) methods—with semiclassical mean-field descriptions of molecular subsystems have proven remarkably successful. Such mixed quantum–classical approaches have demonstrated the ability to reproduce a wide range of strong light–matter phenomena.\cite{taflove2005computational, teixeira2007fdtd, mcmahon2007tailoring, zhao2008methods, sukharev2011numerical, puthumpally2014dipole, sukharev2017optics, you2019nonlinear, sidler2020polaritonic, tancogne2020octopus, sukharev2021second, zhou2024nature} 
A major advantage of FDTD lies in its straightforward extraction of optical response functions, which is essential for spectroscopic analysis. 
Moreover, nonlinear processes involving multiple relevant frequencies can be naturally captured without the prohibitive cost of full quantum simulations. \cite{sukharev2017optics, cui2023comparing}
Finally, for dense molecular layers, dipole–dipole interactions become significant and can give rise to pronounced collective behaviors.  \cite{chuang2021universal, cui2023comparing, cao2025cavity}

At the same time, the classical description of the radiation field cannot account for the quantum character of spontaneous emission and spontaneous Raman scattering (see, e.g. Ref. \citenum{chen2019ehrenfest+} and \citenum{chen2019ehrenfest}). Moreover, Raman signals are second order effects, while a strong driving leads to strong linear background response that hitherto have precluded us from observing clean Raman signals. 
Building on these considerations, the present work investigates the intensity and structure of {\em stimulated} Raman signals from a molecular layer, comparing the stimulated Raman response of such a layer in vacuum to the analogous response  near a single metallic mirror and inside a cavity made of two metallic mirrors. The stimulated Raman signals are obtained by pumping the system with a central driving frequency $\omega_{d}$ and triggering stimulated emissions with a probe short pulse. We obtain the linear absorption and stimulated Raman spectra for molecular models characterized by their free-space Franck-Condon structures, making it possible to examine the effects of the mirror and cavity environment on these structures. Our simulations reveal four key findings:
(i) A cavity consisting of two mirrors further enhances Raman signals as compared to a single flat metal surface by promoting a larger absorption cross-section.
(ii) Formation of cavity polaritons allow more absorption at different energetic regimes. 
(iii) A low-quality cavity introduces extra broadening on the polaritons, leading to richer Raman signatures over a wider range of frequencies.
(iv) The local inhomogeneous dielectric environment can modify the Raman signal profile and exhibit a non-Gaussian/Lorentzian lineshape.

This manuscript is arranged as follows. In section \ref{sec: two}, we present our model scenarios with one layer of 2-level molecules and one mirror/one cavity and how we apply FDTD method and solve the coupled Maxwell-Schrodinger equation to extract stimulated Raman signals.  
In section \ref{sec: three}, we investigate the absorption and stimulated Raman spectra for molecular models characterized by their free space Franck-Condon structures. 
The effect of the driving frequency on the calculated stimulated Raman spectrum is also studied.
We conclude in section \ref{sec: four} where we summarize our findings and discuss the importance of the local dielectric environment on the observed spectra.

\section{Model and Methods \label{sec: two}}
\subsection{Structure of the systems}
\begin{figure}[!t]
    \includegraphics[width=15cm]{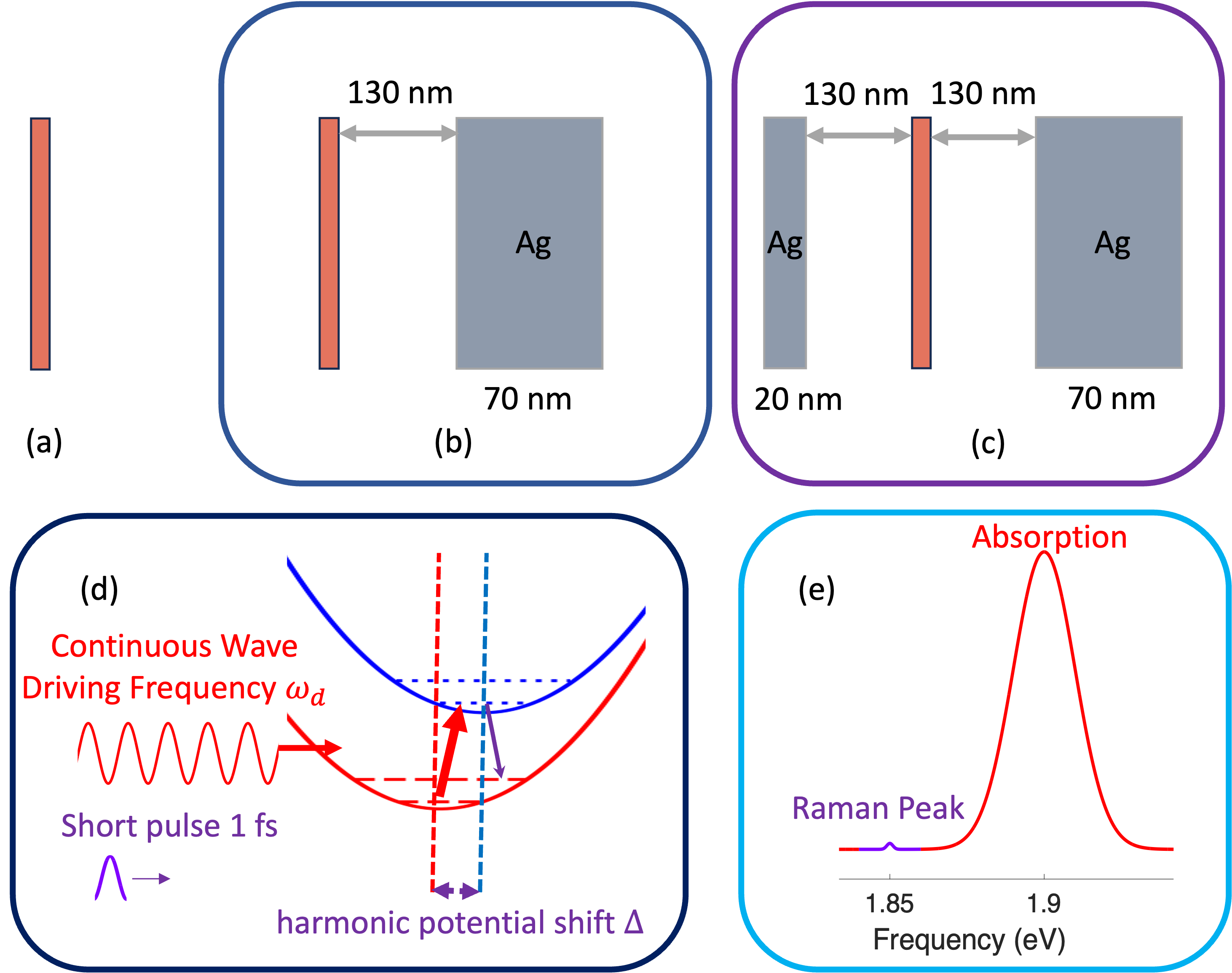}
    \caption{Sketches of three model scenarios: (a) a bare molecular layer, (b) a molecular layer with one $70$ nm silver mirror and (c) a molecular layer inside a cavity formed by $70 + 20$ nm mirrors. The $70$ nm mirror ensures no transmission leaking to the left side. The $20$ nm mirror is thin such that non-resonant light (short pulse and Raman peaks) can efficiently penetrate through. (d) Relevant energy levels inside the molecular layer. The calculation is performed with one continuous wave that promote population onto the excited electronic state $\ket{e0}$ from the ground state $\ket{g0}$. After $3$ ps delay, a short pulse stimulates the emission from $\ket{e0}$ to $\ket{g1}$. (e) A representative output in absorption spectra.}
    \label{fig: systemsketch}
\end{figure}
In this work, we consider three scenarios.
First, as shown in Fig \ref{fig: systemsketch}(a), we imagine a single layer of molecules in vacuum. Second, as shown in Fig \ref{fig: systemsketch} (b), we add a thick silver mirror ($70$ nm) on the righthand side of the same molecular layer as in (a). Last, as shown in Fig \ref{fig: systemsketch} (c), we place a thin silver mirror ($20$ nm) on the left side of the molecular layer and thus a relatively low-quality cavity is formed.
Note that the stimulated Raman signal is often off-resonance with the cavity polaritons and therefore relatively weak. The use of a thin mirror ensures large enough Raman signals to leak out and get collected on the left side.
In Fig. \ref{fig: systemsketch} (d), we sketch the relevant molecular energy diagram. In our numerical implementation, each spatial grid point represents a molecule, modeled by one harmonic vibrational mode $\{\ket{0}, \ket{1}, ...\}$ ($\hbar\omega_{vib} = 0.05$ eV) and two electronic potential surfaces (red and blue parabolas, $\{\ket{g}, \ket{e}\}$). The ground (red parabola) and excited (blue parabola) electronic states are vertically separated by $1.9$ eV and their horizontal shift is denoted by the parameter $\Delta$.

The stimulated Raman signal is obtained by (i) driving the system with a continuous wave (red wave) with central frequency $\omega_{d}$ until a steady state is reached on the electronic excited state (the red arrow) and (ii) upon reaching a steady state, a short pulse ($1$ fs long) triggers emissions (the purple pulse and arrow).
Finally, Fig \ref{fig: systemsketch}(e) displays a sample absorption spectrum with a bare absorption peak (red) and a small Raman peak (purple). The driving frequency is at $\hbar\omega_{d} = 1.90$ eV and thus the primary Raman peak appears at $\hbar\omega_{d} - \hbar\omega_{vib} = 1.85$ eV. Note that fig \ref{fig: systemsketch}(e) is an ideal schematic figure purely for illustrative purposes.

\subsection{Maxwell-Schrodinger equations\label{subsec: maxwellequations}}
We simulate the electromagnetic field by a set of one-dimensional Maxwell equations as
\begin{align}
    \frac{\partial {\cal B}_{y}}{\partial t} &= -\frac{\partial {\cal E}_{x}}{\partial{z}}\label{eq: mag}\\
    \epsilon_{0}\frac{\partial {\cal E}_{x}}{\partial t} &= -\frac{\partial {\cal B}_{y}}{\mu_{0}\partial{z}} - J_{x}\label{eq: elec}
\end{align}
Here, $\epsilon_{0}$ is the vacuum permittivity, $\mu_{0}$ is the vacuum permeability, $z$ is the longitudinal coordinate that is perpendicular to the mirror and molecular slab. $J_{x}(z)$ is the polarization current on grid point ($z$) defined as
\begin{align}
    J_{x}(z) = \frac{dP_{x}(z)}{dt} = \varrho\frac{d(\bra{\psi(z, t)}\hat{\mu}_{x}(z)\ket{\psi(z, t)})}{dt}
    \label{eq: current}
\end{align}
Here, $\varrho$ is the number density, $\mu_{x}$ is a matrix of transition dipole moments between electronic states along $x$ direction.

Within the silver mirror, we define a macroscopic polarization $P_{x}$ following the standard Drude model:
\begin{align}
\frac{\partial^2 P_{x}}{\partial t^2} + \gamma \frac{\partial P_{x}}{\partial t} = \epsilon_{0}\Omega_{p}^{2}{\cal E}_{x}
\label{eq: silver}
\end{align}
Here, the damping rate and plasma frequency for silver are defined as $\gamma = 0.243$ eV and $\Omega_{p} = 8.90$ eV.
For details regarding the FDTD solver for Maxwell equations with different dielectric constants, please see \citenum{taflove2005computational} and \citenum{sukharev2023efficient}.

Finally, the evolution of the wavefunctions $\ket{\psi_{z_{i}}(r, t)}$ inside the molecular layer ($\vert z_{i}\vert< 10$ nm) is described by Schrodinger equation under a mean-field approximation. In other words, there is one wavefunction at each spatial grid point within the molecular layer, the equation of motion is
\begin{align}
    i\hbar\frac{d}{dt}\ket{\psi_{z_{i}}(r, t)} &= \hat{H}(t)\ket{\psi_{z_{i}}(r, t)}.
\end{align}
The time-dependent Hamiltonian 
\begin{align}
\hat{H} =& \Bigl(-\frac{\hbar^{2}}{2m}\frac{\partial^{2}}{\partial r^2} + \frac{1}{2}m\omega_{vib}^2\hat{r}^2\Bigr)\ketbra{g}{g}+\Bigl(-\frac{\hbar^{2}}{2m}\frac{\partial^{2}}{\partial r^2} + \frac{1}{2}m\omega_{vib}^2(\hat{r}-\Delta\hat{I}_{r})^2+\hbar\omega_{ge}\Bigr)\ketbra{e}{e} 
\nonumber\\
&+ \mu^{ge}_{x} {\cal E}_{x}(t)(\ketbra{g}{e} + \ketbra{e}{g})
\end{align}
includes the spatially local time-dependent electric field as the coupling between adjacent electronic states, i.e. $\bra{g}\hat{H}\ket{e} = \mu_{x}^{ge}{\cal E}_{x}(t)$. 
Here, $\omega_{vib} = 0.05 \text{eV}$, $\Delta$ is the harmonic potential shift that determines the Franck-Condon spectral structure. All wavefunctions inside the molecular layer are initialized on the ground vibronic state ($\ket{g0}$). 
Moreover, to ensure the numerical stability of the FDTD calculation, we introduce a phenomenological relaxation time $T_{1} = 1 \text{ ps}$ and a dephasing time $T_{2} = 10 \text{ ps}$, which is motivated by the inherent lifetime of the excited state. For numerical efficiency, instead of using a set of Liouville equations, we apply a non-Hermitian wave packet approximation as in Refs \citenum{charron2013non}, \citenum{puthumpally2016non}, and \citenum{sukharev2017molecular}.
The spatial grid in our simulations is a one-dimensional array from $-1000$ nm to $1000$ nm, with $2000$ grid points.
Near the left edge of the simulation cell at $z_{inc} = -950$ nm, we drive one grid point with an electric field that is the sum of a continuous wave and a delayed $1$ fs short pulse. This driving generates an EM field traveling in both directions. In the rightward direction, the generated EM field reaches the nanolayers at normal incidence.
In the leftward direction, the generated EM field is absorbed by a convolutional perfectly matched layer (CPML), a boundary condition that ensures no reflections back into the system. 
\cite{taflove2005computational}

At each time step, we record the EM fields at the two sides which are effectively transmission and reflection signals. 
We perform a Fourier transformation and normalize by the incident amplitudes to find the transmission ($\mathbf{T}$) and reflection ($\mathbf{R}$) spectra. 
The absorption spectra are obtained by the normalization condition as $\mathbf{A} = 1 -\mathbf{T} - \mathbf{R}$. 
We compare absorption spectra with vs without external driving field and calculate the difference to extract Raman peaks.

To obtain the absorption spectrum in the presence of an external driving (pump) field, it is necessary to isolate the probe-induced response from the total electromagnetic (EM) field (which contains contributions from both the pump and probe). To this end, we perform two simulations in parallel: one with both pump and probe fields, and one with only the pump field. The probe-induced electric field is then obtained by subtracting the two fields at each time step:
\begin{align}
{\cal E}(t;\,\text{probe with pump}) = {\cal E}(t;\,\text{pump+probe}) - {\cal E}(t;\,\text{pump only}) \, .
\end{align}
Before the probe pulse is introduced (at $t = 3\,\mathrm{ps}$), this difference is strictly zero, ensuring that only probe-induced dynamics are captured.

The corresponding Poynting vector is computed as
\begin{align}
{\cal S}(t; \text{probe with pump}) = \frac{{\cal E}(t; \text{probe with pump}) \times {\cal B}(t; \text{probe with pump})}{\mu_0} 
\end{align}

Finally, the signal is normalized by the incident probe intensity, and a Fourier transform of the Poynting vector is performed to obtain the absorption spectrum under continuous-wave (CW) pumping. This result is compared with the standard probe-only spectrum (i.e., without pumping) to extract the nonlinear response induced by the pump field.
See appendix \ref{apdx: para} for details of our parameter choice. 

\section{Results \label{sec: three}}
In this section, we present our FDTD simulation results from two perspectives. First, we present how the linear absorption signal (which in turn determines the Raman signals strength) depends on the harmonic potential shift $\Delta$, because the absorption strength determines the Raman signals strength.
Second, we study the stimulated Raman signals strength for varying the driving frequency $\omega_{d}$ and harmonic potential shift $\Delta$.

\subsection{Linear absorption intensity for varying harmonic potential shift}
\begin{figure}[t!]
\includegraphics[width=15cm]{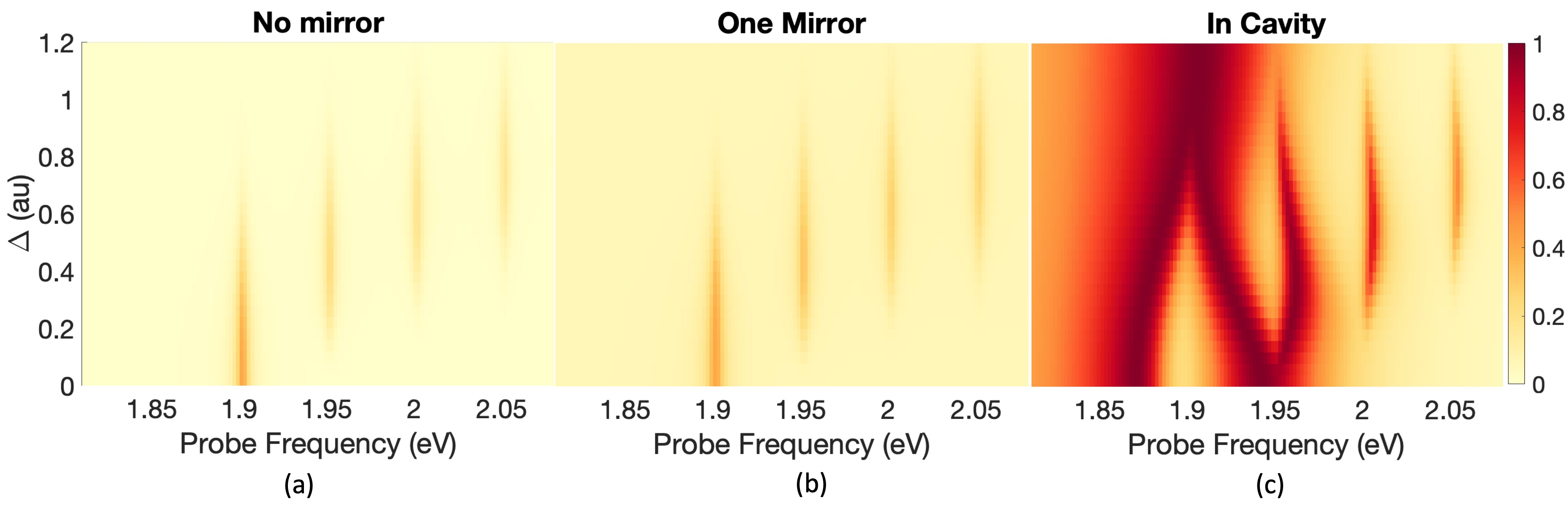}
\caption{Linear absorption spectra with varying harmonic potential shift $\Delta$ for (a) no mirror, (b) one mirror and (c) inside a cavity. The molecular layer is positioned at distance 130 nm from the miror(s) as indicated in Fig \ref{fig: systemsketch}. There is relatively smaller absorption for the no mirror scenario, because we choose the mirror-molecule distance in scenarios (b) and (c) such that constructive interference between incident and reflected wave enhances the EM field amplitude near the molecular layer.}
\label{fig: lin_abs_spec}
\end{figure}
\begin{figure}[t!]
\includegraphics[width=\textwidth]{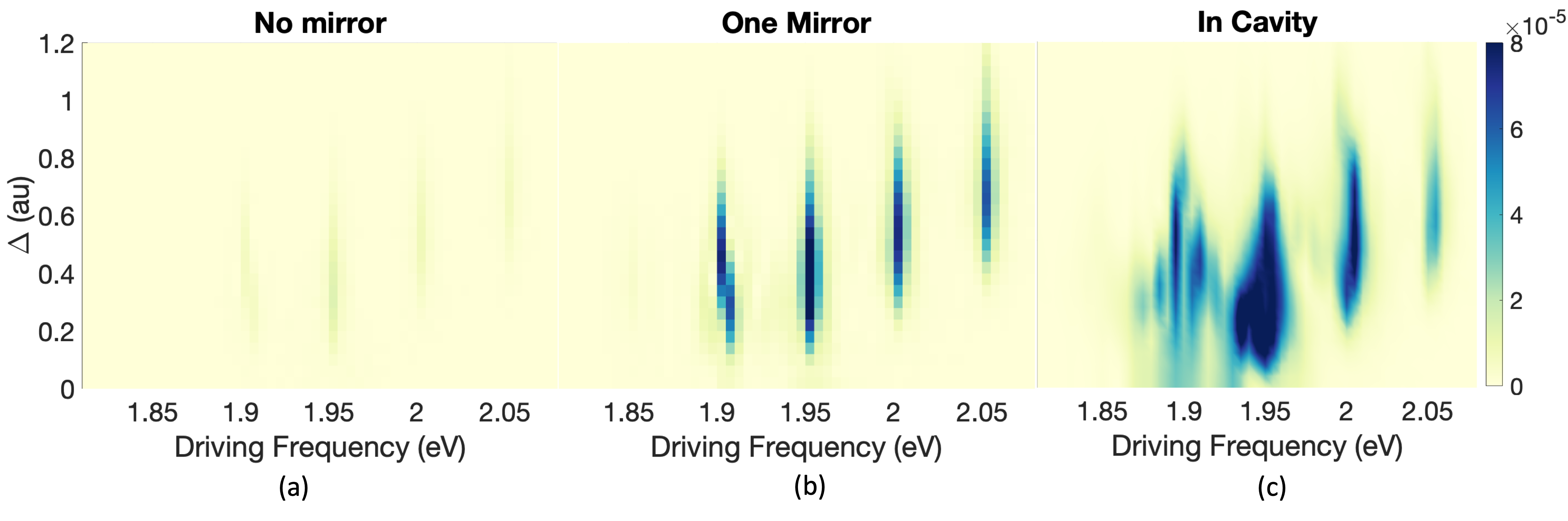}
\caption{Integrated Raman signal strength for varying driving frequencies ($\omega_{d}$) and harmonic potentials shift ($\Delta$) for the three scenarios shown in Fig \ref{fig: systemsketch}(a)-(c). In all cases, the integrated Raman signal strength is found to be proportional to the linear absorption at frequency $\omega_{d}$ in Fig \ref{fig: lin_abs_spec}. Inside a cavity, however, the the signal has a rich signature resulting from its dependence on the local dielectric response  at any given driving frequencies $\omega_{d}$. Thereby, we observe several thin peaks (following the intrinsic long lifetime of the bare molecular excited state $\ket{e0}$ and the external pumping field) spanning a large area (following the short lifetime of the polaritonic states.)}
\label{fig: Raman_signal}
\end{figure}
We begin by investigating the linear absorption intensity as a function of the harmonic potential shift $\Delta$ between the ground and excited electronic states within the molecular layer. 
Here, a linear absorption spectrum is obtained by sending a short pulse into the system in the absence of earlier driving.
As shown in Fig \ref{fig: lin_abs_spec}(a), we observe four peaks (within the frequency range) corresponding to transitions from $\ket{g0}$ to $\ket{e0}$ - $\ket{e3}$ with energies $1.90, 1.95, 2.00, 2.05$ eV (within the focused energy range).  
Moreover, the peaks reach their maxima at $\Delta \approx 0, 0.4, 0.58, 0.71$ respectively in accordance with the maximal overlap between the vibrational wavefunctions, that is, the Franck-Condon envelope.
Next, in Fig \ref{fig: lin_abs_spec}(b), with a silver mirror placed on the right hand side of the molecular layer, a constructive interference between incident and reflected field enhances linear absorption strength (the peaks appear as dark red). 
Last, in Fig \ref{fig: lin_abs_spec}(c), the molecular layer is placed between two mirrors with the fundamental cavity model taken in resonance with the molecular  $\ket{g0}\rightarrow\ket{e0}$ ($1.9$ eV) transition. We see polaritonic peaks at $\Delta \approx 0$, and a shrinking of the Rabi splitting as $\Delta$ increases and the overlap decreases. Note that the cavity environment further enhances the local field and consequently the absorption strength. Note also that the finite quality factor of the silver-mirrored cavity yields very broad polaritonic peaks.

\subsection{Raman intensity}
Now, we turn on the driving field with varying central angular frequency $\omega_{d}$. 
The integrated stimulated Raman signals (obtained from the difference between the emitted intensity with and without the driving field and integrated over $\hbar\omega_{out} = \hbar\omega_{d}-\hbar\omega_{vib} \pm 0.01$ eV) are shown in Fig \ref{fig: Raman_signal} for different values of $\omega_{d}$ and $\Delta$ for three model systems shown in Fig \ref{fig: systemsketch} (a)-(c).
\footnote{The detailed shape of these Raman signals as functions of the outgoing frequency are strongly dependent on the local dielectric environment and may not look like a perfect Gaussian-Lorentzian shape as the purple peak in Fig \ref{fig: systemsketch} (e)}

As shown by comparing Fig \ref{fig: lin_abs_spec} and Fig \ref{fig: Raman_signal} (a) and (b), the Raman signal intensity increases with the absorption strength, and in fact we find that these observables are proportional to each other. Furthermore, as expected, the Raman signal intensity vanishes at $\Delta = 0$.
By contrast, as shown in Fig \ref{fig: Raman_signal} (c), a cavity can change some quantitative features of the these Raman signals. 
First, the broadening due to the cavity in the linear absorption spectra in Fig \ref{fig: lin_abs_spec}(c) leads to broader Raman signals. 
Second, when the cavity (model (c)) is driven near the two polariton branches, the Raman signals strength show an interference pattern that has a structure associated with the stimulating CW linewidth modulating the polariton signals.
In particular, small changes in effective light-matter couplings (associated with its dependence on $\Delta$) and the driving frequencies ($\hbar\omega_{dr}$) lead to phase mismatching and rematching, leading to an oscillatory behavior of the Raman signal strength as a function of the driving frequency $\omega_{d}$, as seen in Fig \ref{fig: Raman_signal} (c).
Last, we observe a strong signal when the upper polariton coincides with the $\ket{g0}\rightarrow\ket{e1}$ transition  at 1.95 eV.

One more interesting, and potentially puzzling, observation is the appearance of a Raman signal at $\Delta = 0$ between the two polariton branches at $1.85$ and $1.95$ eV. 
Such a signal should vanish in steady-state scattering because for $\Delta = 0$, there is no coupling between different vibrational states on different surfaces. In a transient pump-probe process, during the time-lag between populating the excited electronic state and the subsequent stimulated emission induced by the short pulse, the depletion of ground state population leads to a Rabi contraction (also seen in Fig \ref{fig: Raman_signal}).\cite{son2022energy, stemo2024ultrafast, jin2025excited, chen2025ultrafast} In Appendix \ref{apdx: ramanvsrabi}, we present the direct comparison between a Rabi contraction signal and a Raman signal. 
\begin{figure}[t!]
\includegraphics[width=\textwidth]{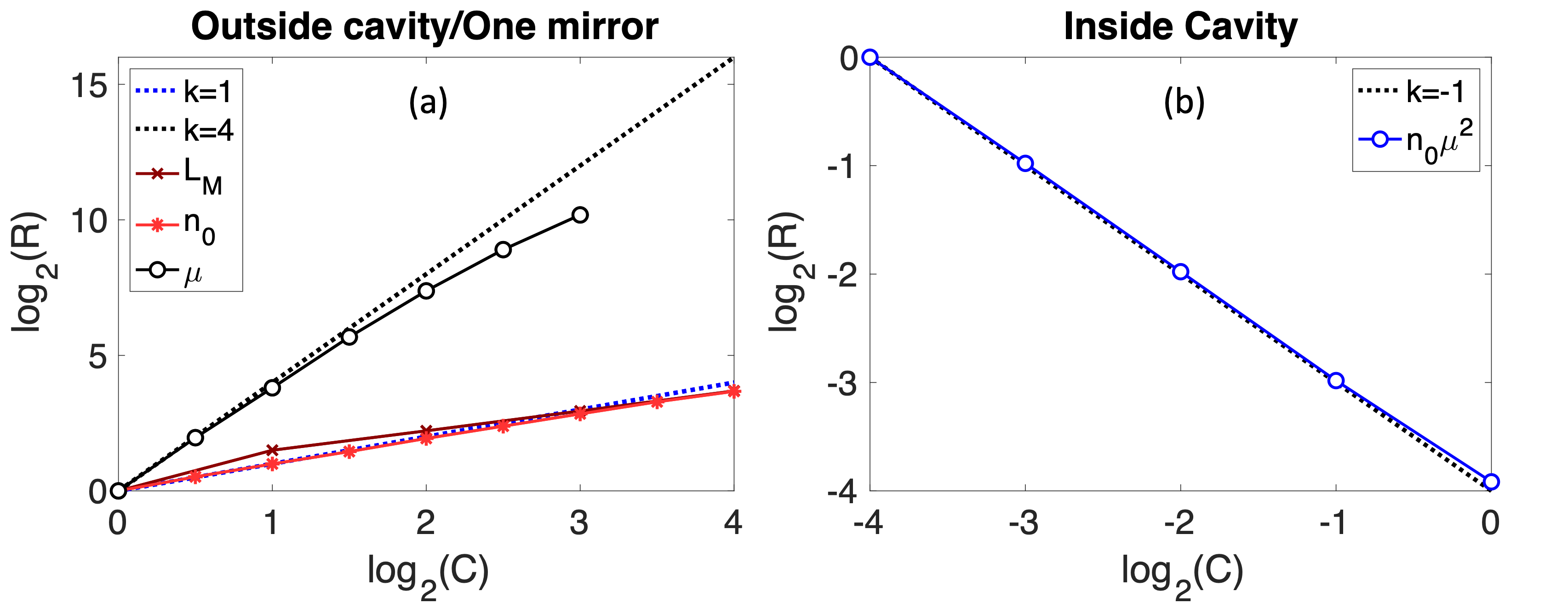}
\caption{Raman intensity ($R$) dependence on the transition dipole moment $\mu$, molecular density (per unit volume) $n_{0}$, and layer thickness $L_{M}$ for (a) outside a cavity and (b) inside a cavity. On the vertical axis, we plot the log of the Raman peak (non-integrated) height.  The three dotted lines in both subfigures serve as references, corresponding to power-law slopes $k = 1$, $k = 4$, and $k = -1$, respectively.
As shown by the red lines in (a), the Raman intensity scales linearly with both $L_{M}$ and $n_{0}$, indicating the absence of collective effects. Note that $L_{M}$ and $n_{0}$ both characterize the number of molecules in the system and have identical effects on the Raman signals. The black line demonstrates the quartic dependence $R \propto \mu^{4}$, consistent with the two-photon nature of the Raman process.
For a meaningful quantitative comparison of Raman signals inside the cavity, we impose a constant Rabi splitting, $\Omega_{R} \sim \sqrt{\mu^{2} n_{0} L_{M}} \equiv \text{const}$. Under this constraint, the Raman signal exhibits an apparent inverse dependence on the prefactor, $R \propto 1/C$. See eq \ref{eq: C} for a definition of $C$. }
\label{fig: Raman_signal_dep}
\end{figure}

Next, we investigate how the Raman signal intensity $R$ depends on key physical parameters: transition dipole moment $\mu$, molecular number density $n_{0}$, and molecular layer thickness $L_{M}$. To extract the relevant scaling laws, we introduce a common prefactors $C$, apply to all quantities, and then analyze the Raman signals to extract the corresponding power-law dependencies.
As shown in Fig.~\ref{fig: Raman_signal_dep}(a), in the absence of a cavity, the normalized Raman signal $R$ scales linearly with the prefactor $C$ for both $n_{0}$ and $L_{M}$. In addition, the Raman signal exhibits the expected $\mu^{4}$ dependence, consistent with the two-photon nature of the Raman process.

As observed in Figs.~\ref{fig: lin_abs_spec} and \ref{fig: Raman_signal}, the stimulated Raman signal strength is positively correlated with the linear absorption. Therefore, for scenario (c) in Fig.~\ref{fig: systemsketch}, in order to isolate cavity-induced effects, we constrain the linear absorption spectrum to remain unchanged while varying system parameters. This is achieved by simultaneously scaling $\mu$ and $n_{0}$ as
\begin{equation}
\mu \rightarrow \mu / \sqrt{C}, \quad n_{0} \rightarrow n_{0} C,
\label{eq: C}
\end{equation}
which keeps the Rabi splitting $\Omega_{R} \sim \sqrt{\mu^{2} n_{0}}$ constant.

Under this constraint, Fig.~\ref{fig: Raman_signal_dep}(b) shows that the Raman signal inside the cavity scales inversely with the prefactor $C$ associated with the molecular density $n_{0}$. This behavior is fully consistent with the scaling relation $R \propto \mu^{4} n_{0}$, confirming that the apparent inverse dependence arises from the imposed constraint that preserves the linear absorption (and hence the light--matter coupling strength).
\section{Conclusion and Outlook\label{sec: four}}
We have investigated the effect on the stimulated Raman signal induced by a single mirror and a low-quality cavity by solving the coupled Maxwell--Schr\"odinger equations using the FDTD technique. We present the integrated Raman signal as a function of the external driving frequency and the harmonic potential shift.
In contrast to a bare molecular layer, we find that, in the single-mirror scenario (b), the enhancement of the Raman signal strength is primarily proportional to the corresponding enhancement in linear absorption induced by the mirror. Furthermore, cavity broadening allows more electromagnetic fields to enter the cavity, leading to stronger Raman peaks near both the lower and upper polariton branches.
The interplay between cavity-induced broadening and the external pumping broadening produces detailed features in the Raman signal as a function of the driving frequency $\omega_{d}$. These features arise from interference of electromagnetic
fields within the local dielectric environment, which occurs when the cavity-induced broadening overlaps with the external pumping broadening.
At large Rabi splitting, we observe that the upper polariton interferes with a higher vibronic state. All these phenomena are captured by our FDTD simulation, which explicitly accounts for the full dielectric environment.

Our work highlights the fact that, if we want to quantitatively understand experimental results, it is important to self-consistently 
treat the molecular quantum subsystem and the local dielectric environment.
For example, a recent experiment has associated differences in the Raman signal as found in the presence of a cavity environment with a possible change of the Franck-Condon envelope on the molecular potential energy surface.\cite{alam2025quantification}
In the system studied here, we observe a strong cavity effect on a Raman signal that arises not from changes in the molecular potential energy surface but rather from a combination of geometrical and optical factors, a distinction that will need to be investigated further in the future as different effects are teased apart.
More generally, looking forward, a sub-wavelength molecular layer opens plenty opportunities for engineering optical devices with unique collective responses. One future direction of this research will be to extend the methodology used here to investigate more coherent nonlinear processes such as two-dimensional spectroscopy in the presence of optical nanomaterials, especially in the strong light-matter coupling regime. 

\section*{Acknowledgments}
This work has been supported by the European Research Council (ERC-2024-SyG-101167294; UnMySt) (AN). 
MS research is supported by the Office of Naval Research, Grant No. N000142512090 and the Air Force Office of Scientific Research under Grant No. FA9550-25-1-0096.
This work was also supported by the U.S. Department of Energy, Office of Science, Office of Basic Energy Sciences, under Award No. DOE-SC0025393 (JES)

\appendix
\section{Two competing second order effects: Rabi contraction vs Raman signals \label{apdx: ramanvsrabi}}
\begin{figure}[t!]
\includegraphics[width=0.5\textwidth ]{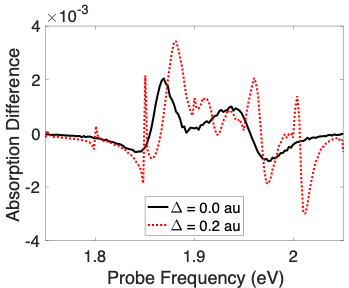}
\caption{The difference in linear absorption due to external pumping for $\Delta = 0.0$ (black solid line) vs $\Delta = 0.2$ (red dotted line); these signals are second order in the external electric field. 
For $\Delta = 0.0$, we observe a smooth line with two differential peak structures (corresponding to the lower and upper polariton peaks). For $\Delta = 0.2$, we observe narrower Raman peaks including the primary one at frequency $\hbar\omega = 1.85 \text{eV}$.}
\label{fig: RamanvsRSC}
\end{figure}

In this appendix, we discuss the non-zero signals found underneath the two polariton branches near $\Delta \approx 0$ in Fig \ref{fig: Raman_signal}.
In Fig.~\ref{fig: RamanvsRSC}, 
the $y$-axis represents the difference in linear absorption for model (c), comparing the case with external continuous-wave pumping ($\hbar\omega_{dr} = 1.9 \text{ eV}$) to that without pumping. This procedure is the same procedure as in Fig. \ref{fig: Raman_signal} (effectively removing the linear response associated with the short pulse and isolating the nonlinear contribution) but now we Fourier transform the problem signal and report data as a function of the resulting probe frequency (which should not be confused the x-axis in Fig. \ref{fig: Raman_signal} which is the {\em driving} frequency).
For the black curve ($\Delta = 0.0$), two differential peaks are observed near the lower and upper polariton energies. 
Each peak exhibits both negative and positive components, with the positive regions located between the two polariton resonances. 
These differential peaks indicate that, under pumping, the polariton peaks shift toward each other, a phenomenon known as Rabi contraction \cite{stemo2024ultrafast, jin2025excited}.
This behavior becomes far more complicated when we introduce a non-zero displacement $\Delta$, and we assign the most prominent Raman peak to be that which occurs at $\hbar\omega = 1.85 \text{ eV}$.
For this driving frequency, for all values of $\Delta$,
we integrate the probe frequency data between $1.84 \text{ eV}$ and $1.86 \text{ eV}$ to obtain the data points in Fig \ref{fig: Raman_signal} (c).
We have numerically verified that the relative amplitudes of the differential peaks and of the Raman peaks are insensitive to the pumping intensity, suggesting that both contributions are of the same order (though not explicitly shown here). 

\section{Parameter\label{apdx: para}}
\begin{table}[ht!]
  \begin{threeparttable}
   \caption[]{Parameters for Maxwell-Bloch simulation.}
   \centering
   \begin{tabular}{cc}
     \midrule 
     Name & Value\tnote{$\dagger$}
    \\\hline
    Molecules electronic resonant frequency $\hbar\omega_{ge}$ & $1.9$ eV
    \\
    Molecules resonant wavelength $\lambda_{M}$ & $652$ nm
    \\
    Molecules vibrational mode frequency $\omega_{vib}$ & $0.05$ eV
    \\
    Molecules relaxation timescale $T_{1}$ & $1$ ps
    \\
    Molecules dephasing timescale $T_{2}$ & $10$ ps
    \\
    Thickness of the silver mirror & $70$ nm \& $20$ nm
    \\
    Distance between silver mirror and molecular layer & $130$ nm
    \\
    Drude model for silver & $\Omega_{d} = 8.90$eV
    \\
    & $\Gamma_{d} = 0.243$ eV
    \\
    Transition dipole moment & $\mu_{x}= 5$ Debye
    \\
    Simulation grid size (1D)& $2000$ nm
    \\
    Number of grid points for EM field & $2000$
    \\
    Harmonic potential grid size (1D) & 12 au
    \\
    Number of grid points for wavefunction & $1200$
    \\
    Number of Wavefunctions & $10$
    \\
    Time step (dt) & $dx/2/c_{0}$
    \\
    CPML grid points & 19
    \\
    \midrule
     \end{tabular}
  \end{threeparttable}
\end{table}
\section*{References}
\bibliography{reference}

@article{taflove2005computational,
  title={Computational electromagnetics: the finite-difference time-domain method},
  author={Taflove, Allen and Hagness, Susan C and Piket-May, Melinda},
  journal={The Electrical Engineering Handbook},
  volume={3},
  number={629-670},
  pages={15},
  year={2005},
  publisher={Elsevier Amsterdam, The Netherlands}
}

@article{sukharev2017optics,
  title={Optics of exciton-plasmon nanomaterials},
  author={Sukharev, Maxim and Nitzan, Abraham},
  journal={Journal of Physics: Condensed Matter},
  volume={29},
  number={44},
  pages={443003},
  year={2017},
  publisher={IOP Publishing}
}

@article{bayer2001coupling,
  title={Coupling and entangling of quantum states in quantum dot molecules},
  author={Bayer, M and Hawrylak, Pawel and Hinzer, K and Fafard, S and Korkusinski, Marek and Wasilewski, ZR and Stern, O and Forchel, A},
  journal={Science},
  volume={291},
  number={5503},
  pages={451--453},
  year={2001},
  publisher={American Association for the Advancement of Science}
}

@article{walther2006cavity,
  title={Cavity quantum electrodynamics},
  author={Walther, Herbert and Varcoe, Benjamin TH and Englert, Berthold-Georg and Becker, Thomas},
  journal={Reports on Progress in Physics},
  volume={69},
  number={5},
  pages={1325},
  year={2006},
  publisher={IOP Publishing}
}

@article{hutchison2012modifying,
  title={Modifying chemical landscapes by coupling to vacuum fields},
  author={Hutchison, James A and Schwartz, Tal and Genet, Cyriaque and Devaux, Elo{\"\i}se and Ebbesen, Thomas W},
  journal={Angewandte Chemie International Edition},
  volume={51},
  number={7},
  pages={1592--1596},
  year={2012},
  publisher={Wiley Online Library}
}

@article{ebbesen2016hybrid,
  title={Hybrid light--matter states in a molecular and material science perspective},
  author={Ebbesen, Thomas W},
  journal={Accounts of chemical research},
  volume={49},
  number={11},
  pages={2403--2412},
  year={2016},
  publisher={ACS Publications}
}

@book{kavokin2017microcavities,
  title={Microcavities},
  author={Kavokin, Alexey V and Baumberg, Jeremy J and Malpuech, Guillaume and Laussy, Fabrice P},
  volume={21},
  year={2017},
  publisher={Oxford university press}
}

@article{ribeiro2018polariton,
  title={Polariton chemistry: controlling molecular dynamics with optical cavities},
  author={Ribeiro, Raphael F and Mart{\'\i}nez-Mart{\'\i}nez, Luis A and Du, Matthew and Campos-Gonzalez-Angulo, Jorge and Yuen-Zhou, Joel},
  journal={Chemical science},
  volume={9},
  number={30},
  pages={6325--6339},
  year={2018},
  publisher={Royal Society of Chemistry}
}

@article{frisk2019ultrastrong,
  title={Ultrastrong coupling between light and matter},
  author={Frisk Kockum, Anton and Miranowicz, Adam and De Liberato, Simone and Savasta, Salvatore and Nori, Franco},
  journal={Nature Reviews Physics},
  volume={1},
  number={1},
  pages={19--40},
  year={2019},
  publisher={Nature Publishing Group}
}

@article{hertzog2019strong,
  title={Strong light--matter interactions: a new direction within chemistry},
  author={Hertzog, Manuel and Wang, Mao and Mony, J{\"u}rgen and B{\"o}rjesson, Karl},
  journal={Chemical Society Reviews},
  volume={48},
  number={3},
  pages={937--961},
  year={2019},
  publisher={Royal Society of Chemistry}
}

@article{garcia2021manipulating,
  title={Manipulating matter by strong coupling to vacuum fields},
  author={Garcia-Vidal, Francisco J and Ciuti, Cristiano and Ebbesen, Thomas W},
  journal={Science},
  volume={373},
  number={6551},
  pages={eabd0336},
  year={2021},
  publisher={American Association for the Advancement of Science}
}

@article{teixeira2007fdtd,
  title={FDTD/FETD methods: a review on some recent advances and selected applications},
  author={Teixeira, Fernando L},
  journal={Journal of Microwaves, Optoelectronics and Electromagnetic Applications (JMOe)},
  volume={6},
  number={1},
  pages={83--95},
  year={2007}
}

@article{mcmahon2007tailoring,
  title={Tailoring the sensing capabilities of nanohole arrays in gold films with Rayleigh anomaly-surface plasmon polaritons},
  author={McMahon, Jeffrey M and Henzie, Joel and Odom, Teri W and Schatz, George C and Gray, Stephen K},
  journal={Optics express},
  volume={15},
  number={26},
  pages={18119--18129},
  year={2007},
  publisher={Optica Publishing Group}
}

@article{zhao2008methods,
  title={Methods for describing the electromagnetic properties of silver and gold nanoparticles},
  author={Zhao, Jing and Pinchuk, Anatoliy O and McMahon, Jeffrey M and Li, Shuzhou and Ausman, Logan K and Atkinson, Ariel L and Schatz, George C},
  journal={Accounts of chemical research},
  volume={41},
  number={12},
  pages={1710--1720},
  year={2008},
  publisher={ACS Publications}
}

@article{sukharev2011numerical,
  title={Numerical studies of the interaction of an atomic sample with the electromagnetic field in two dimensions},
  author={Sukharev, Maxim and Nitzan, Abraham},
  journal={Physical Review A},
  volume={84},
  number={4},
  pages={043802},
  year={2011},
  publisher={APS}
}

@article{puthumpally2014dipole,
  title={Dipole-induced electromagnetic transparency},
  author={Puthumpally-Joseph, Raiju and Sukharev, Maxim and Atabek, Osman and Charron, Eric},
  journal={Physical review letters},
  volume={113},
  number={16},
  pages={163603},
  year={2014},
  publisher={APS}
}

@article{you2019nonlinear,
  title={Nonlinear optical properties and applications of 2D materials: theoretical and experimental aspects},
  author={You, JW and Bongu, SR and Bao, Q and Panoiu, NC},
  journal={Nanophotonics},
  volume={8},
  number={1},
  pages={63--97},
  year={2019},
  publisher={De Gruyter}
}

@article{sidler2020polaritonic,
  title={Polaritonic chemistry: Collective strong coupling implies strong local modification of chemical properties},
  author={Sidler, Dominik and Schäfer, Christian and Ruggenthaler, Michael and Rubio, Angel},
  journal={The journal of physical chemistry letters},
  volume={12},
  number={1},
  pages={508--516},
  year={2020},
  publisher={ACS Publications}
}

@article{tancogne2020octopus,
  title={Octopus, a computational framework for exploring light-driven phenomena and quantum dynamics in extended and finite systems},
  author={Tancogne-Dejean, Nicolas and Oliveira, Micael JT and Andrade, Xavier and Appel, Heiko and Borca, Carlos H and Le Breton, Guillaume and Buchholz, Florian and Castro, Alberto and Corni, Stefano and Correa, Alfredo A and others},
  journal={The Journal of chemical physics},
  volume={152},
  number={12},
  pages={124119},
  year={2020},
  publisher={AIP Publishing LLC}
}

@article{sukharev2021second,
  title={Second harmonic generation by strongly coupled exciton--plasmons: The role of polaritonic states in nonlinear dynamics},
  author={Sukharev, Maxim and Salomon, Adi and Zyss, Joseph},
  journal={The Journal of Chemical Physics},
  volume={154},
  number={24},
  pages={244701},
  year={2021},
  publisher={AIP Publishing LLC}
}

@article{zhou2024nature,
  title={On the nature of two-photon transitions for a collection of molecules in a Fabry--Perot cavity},
  author={Zhou, Zeyu and Chen, Hsing-Ta and Sukharev, Maxim and Subotnik, Joseph E and Nitzan, Abraham},
  journal={The Journal of Chemical Physics},
  volume={160},
  number={9},
  year={2024},
  publisher={AIP Publishing}
}

@article{yi2025surface,
  title={Surface-enhanced Raman spectroscopy: a half-century historical perspective},
  author={Yi, Jun and You, En-Ming and Hu, Ren and Wu, De-Yin and Liu, Guo-Kun and Yang, Zhi-Lin and Zhang, Hua and Gu, Yu and Wang, Yao-Hui and Wang, Xiang and others},
  journal={Chemical Society Reviews},
  year={2025},
  publisher={Royal Society of Chemistry}
}

@article{frontiera2011surface,
  title={Surface-enhanced femtosecond stimulated Raman spectroscopy},
  author={Frontiera, Renee R and Henry, Anne-Isabelle and Gruenke, Natalie L and Van Duyne, Richard P},
  journal={The journal of physical chemistry letters},
  volume={2},
  number={10},
  pages={1199--1203},
  year={2011},
  publisher={ACS Publications}
}

@article{sharma2012sers,
  title={SERS: Materials, applications, and the future},
  author={Sharma, Bhavya and Frontiera, Renee R and Henry, Anne-Isabelle and Ringe, Emilie and Van Duyne, Richard P},
  journal={Materials today},
  volume={15},
  number={1-2},
  pages={16--25},
  year={2012},
  publisher={Elsevier}
}

@article{haynes2005surface,
author = {Haynes, Christy L. and McFarland, Adam D. and Van Duyne, Richard P.},
title = {Surface-Enhanced Raman Spectroscopy},
journal = {Analytical Chemistry},
volume = {77},
number = {17},
pages = {338 A-346 A},
year = {2005},
}

@article{stiles2008surface,
  title={Surface-enhanced Raman spectroscopy},
  author={Stiles, Paul L and Dieringer, Jon A and Shah, Nilam C and Van Duyne, Richard P},
  journal={Annu. Rev. Anal. Chem.},
  volume={1},
  pages={601--626},
  year={2008},
  publisher={Annual Reviews}
}

@article{hummer2016cavity,
  title={Cavity-enhanced Raman microscopy of individual carbon nanotubes},
  author={H{\"u}mmer, Thomas and Noe, Jonathan and Hofmann, Matthias S and H{\"a}nsch, Theodor W and H{\"o}gele, Alexander and Hunger, David},
  journal={Nature communications},
  volume={7},
  number={1},
  pages={12155},
  year={2016},
  publisher={Nature Publishing Group UK London}
}

@article{saunders2016cavity,
  title={Cavity-enhanced room-temperature broadband Raman memory},
  author={Saunders, DJ and Munns, JHD and Champion, TFM and Qiu, C and Kaczmarek, KT and Poem, E and Ledingham, PM and Walmsley, IA and Nunn, J},
  journal={Physical review letters},
  volume={116},
  number={9},
  pages={090501},
  year={2016},
  publisher={APS}
}

@article{zhang2018dicke,
  title={Dicke-model simulation via cavity-assisted Raman transitions},
  author={Zhang, Zhiqiang and Lee, Chern Hui and Kumar, Ravi and Arnold, KJ and Masson, Stuart J and Grimsmo, AL and Parkins, AS and Barrett, MD},
  journal={Physical Review A},
  volume={97},
  number={4},
  pages={043858},
  year={2018},
  publisher={APS}
}

@article{sukharev2023efficient,
  title={Efficient parallel strategy for molecular plasmonics--A numerical tool for integrating Maxwell-Schr{\"o}dinger equations in three dimensions},
  author={Sukharev, Maxim},
  journal={Journal of Computational Physics},
  volume={477},
  pages={111920},
  year={2023},
  publisher={Elsevier}
}

@article{chuang2021universal,
  title={Universal scalings in two-dimensional anisotropic dipolar excitonic systems},
  author={Chuang, Chern and Cao, Jianshu},
  journal={Physical Review Letters},
  volume={127},
  number={4},
  pages={047402},
  year={2021},
  publisher={APS}
}

@article{cao2025cavity,
  title={Cavity-induced quantum interference and collective interactions in van der waals systems},
  author={Cao, Jianshu and Pollak, Eli},
  journal={The Journal of Physical Chemistry Letters},
  volume={16},
  number={22},
  pages={5466--5472},
  year={2025},
  publisher={ACS Publications}
}

@article{cui2023comparing,
  title={Comparing semiclassical mean-field and 1-exciton approximations in evaluating optical response under strong light--matter coupling conditions},
  author={Cui, Bingyu and Sukharev, Maxim and Nitzan, Abraham},
  journal={The Journal of chemical physics},
  volume={158},
  number={16},
  year={2023},
  publisher={AIP Publishing}
}

@article{chen2019ehrenfest+,
  title={Ehrenfest+ R dynamics. I. A mixed quantum--classical electrodynamics simulation of spontaneous emission},
  author={Chen, Hsing-Ta and Li, Tao E and Sukharev, Maxim and Nitzan, Abraham and Subotnik, Joseph E},
  journal={The Journal of chemical physics},
  volume={150},
  number={4},
  year={2019},
  publisher={AIP Publishing}
}

@article{chen2019ehrenfest,
  title={Ehrenfest+ R dynamics. II. A semiclassical QED framework for Raman scattering},
  author={Chen, Hsing-Ta and Li, Tao E and Sukharev, Maxim and Nitzan, Abraham and Subotnik, Joseph E},
  journal={The Journal of Chemical Physics},
  volume={150},
  number={4},
  year={2019},
  publisher={AIP Publishing}
  }

@article{alam2025quantification,
  title={Quantification of Nuclear Coordinate Activation on Polaritonic Potential Energy Surfaces},
  author={Alam, Shahzad and Liu, Yicheng and Holmes, Russell J and Frontiera, Renee R},
  journal={arXiv preprint arXiv:2501.06364},
  year={2025}
}

@article{stemo2024ultrafast,
  title={Ultrafast spectroscopy under vibrational strong coupling in diphenylphosphoryl azide},
  author={Stemo, Garrek and Nishiuchi, Joel and Bhakta, Harsh and Mao, Haochuan and Wiesehan, Garret and Xiong, Wei and Katsuki, Hiroyuki},
  journal={The Journal of Physical Chemistry A},
  volume={128},
  number={10},
  pages={1817--1824},
  year={2024},
  publisher={ACS Publications}
}

@article{jin2025excited,
  title={Excited State Dynamics of CO2 Reduction Catalyst under Vibrational Strong Coupling},
  author={Jin, Tao and Gebre, Sara T and Miller, Christopher J and Kubiak, Clifford P and Ribeiro, Raphael F and Lian, Tianquan},
  journal={Journal of the American Chemical Society},
  volume={147},
  number={42},
  pages={38320--38330},
  year={2025},
  publisher={ACS Publications}
}

@article{son2022energy,
  title={Energy cascades in donor-acceptor exciton-polaritons observed by ultrafast two-dimensional white-light spectroscopy},
  author={Son, Minjung and Armstrong, Zachary T and Allen, Ryan T and Dhavamani, Abitha and Arnold, Michael S and Zanni, Martin T},
  journal={Nature Communications},
  volume={13},
  number={1},
  pages={7305},
  year={2022},
  publisher={Nature Publishing Group UK London}
}

@article{chen2025ultrafast,
  title={Ultrafast optical modulation of vibrational strong coupling in ReCl (CO) 3 (2, 2-bipyridine)},
  author={Chen, Liying and McKillop, Alexander M and Fidler, Ashley P and Weichman, Marissa L},
  journal={Nanophotonics},
  volume={14},
  number={27},
  pages={5437--5448},
  year={2025},
  publisher={De Gruyter}
}

@article{charron2013non,
  title={Non-Hermitian wave packet approximation of Bloch optical equations},
  author={Charron, Eric and Sukharev, Maxim},
  journal={The Journal of Chemical Physics},
  volume={138},
  number={2},
  year={2013},
  publisher={AIP Publishing}
}

@article{puthumpally2016non,
  title={Non-Hermitian wave packet approximation for coupled two-level systems in weak and intense fields},
  author={Puthumpally-Joseph, Raiju and Sukharev, Maxim and Charron, Eric},
  journal={The Journal of Chemical Physics},
  volume={144},
  number={15},
  year={2016},
  publisher={AIP Publishing}
}

@article{sukharev2017molecular,
  title={Molecular plasmonics: The role of rovibrational molecular states in exciton-plasmon materials under strong-coupling conditions},
  author={Sukharev, Maxim and Charron, Eric},
  journal={Physical Review B},
  volume={95},
  number={11},
  pages={115406},
  year={2017},
  publisher={APS}
}

@article{li2021cavity,
  title={Cavity frequency-dependent theory for vibrational polariton chemistry},
  author={Li, Xinyang and Mandal, Arkajit and Huo, Pengfei},
  journal={Nature communications},
  volume={12},
  number={1},
  pages={1315},
  year={2021},
  publisher={Nature Publishing Group UK London}
}

@article{boolakee2022light,
  title={Light-field control of real and virtual charge carriers},
  author={Boolakee, Tobias and Heide, Christian and Garz{\'o}n-Ram{\'\i}rez, Antonio and Weber, Heiko B and Franco, Ignacio and Hommelhoff, Peter},
  journal={Nature},
  volume={605},
  number={7909},
  pages={251--255},
  year={2022},
  publisher={Nature Publishing Group UK London}
}

@article{sweeney2014cavity,
  title={Cavity-stimulated Raman emission from a single quantum dot spin},
  author={Sweeney, Timothy M and Carter, Samuel G and Bracker, Allan S and Kim, Mijin and Kim, Chul Soo and Yang, Lily and Vora, Patrick M and Brereton, Peter G and Cleveland, Erin R and Gammon, Daniel},
  journal={Nature Photonics},
  volume={8},
  number={6},
  pages={442--447},
  year={2014},
  publisher={Nature Publishing Group UK London}
}

@article{hu2022tuning,
  title={Tuning and enhancing quantum coherence time scales in molecules via light-matter hybridization},
  author={Hu, Wenxiang and Gustin, Ignacio and Krauss, Todd D and Franco, Ignacio},
  journal={The Journal of Physical Chemistry Letters},
  volume={13},
  number={49},
  pages={11503--11511},
  year={2022},
  publisher={ACS Publications}
}
\end{document}